\documentclass[runningheads]{llncs}
\usepackage[T1]{fontenc}
\usepackage{graphicx}
\usepackage{xcolor}
\usepackage{amsmath}
\begin{document}
\title{Mamba Outpaces Reformer in Stock Prediction with Sentiments from Top Ten LLMs}
\titlerunning{Mamba/Reformer with Semantics Score From Top Ten LLMS }
%
\author{Lokesh Antony Kadiyala \and
Amir Mirzaeinia
}
\authorrunning{L.A. Kadiyala et al.}
%
\institute{University of North Texas USA, 
 Discovery Park, 3940 N Elm St, Denton, TX 76207
\email{lokeshantonykadiyala@my.unt.edu}\\
\email{amir.mirzaeinia@.unt.edu}
}
%
\maketitle              
\begin{abstract}
The stock market is extremely difficult to predict in the short term due to high market volatility, changes caused by news, and the non-linear nature of the financial time series. This research proposes a novel framework for improving minute-level prediction accuracy using semantic sentiment scores from ten different large language models (LLMs) combined with minute interval intraday stock price data. We systematically constructed a time-aligned dataset of AAPL news articles and 1-minute Apple Inc. (AAPL) stock prices for the dates of April 4 to May 2, 2025. The sentiment analysis was achieved using the DeepSeek-V3, GPT variants, LLaMA, Claude, Gemini, Qwen, and Mistral models through their APIs. Each article obtained sentiment scores from all ten LLMs, which were scaled to a [0, 1] range and combined with prices and technical indicators like RSI, ROC, and Bollinger Band Width. 
Two state-of-the-art such as Reformer and Mamba were trained separately on the dataset using the sentiment scores produced by each LLM as input. Hyper parameters were optimized by means of Optuna and were evaluated through a 3-day evaluation period. Reformer had mean squared error (MSE) or the evaluation metrics, and it should be noted that Mamba performed not only faster but also better than Reformer for every LLM across the 10 LLMs tested. Mamba performed best with LLaMA 3.3–70B, with the lowest error of 0.137. While Reformer could capture broader trends within the data, the model appeared to over smooth sudden changes by the LLMs. This study highlights the potential of integrating LLM-based semantic analysis paired with efficient temporal modeling to enhance real-time financial forecasting.

\end{abstract}
%
%
\section{Introduction}

Forecasting stock prices is a challenging problem because of high volatility, non-stationarity of price behavior, and the non-linear relationships between influencing variables. Price trends are driven not only by Traditional indicators like earnings reports and macroeconomic conditions, but also unstructured data sources like media headlines, social sentiment, and investor psychology.
Recent advancements in natural language processing (NLP) have made large language models (LLMs) such as GPT, Claude, LLaMA, and DeepSeek capable of state-of-the-art performance in contextual understanding. Other sentiment analysis methods prefer multi-condition filters from static lists of words and shallow classifiers. LLMs have billions of parameters for comprehending context, meaning, emotion, tone, and intent. When they derive sentiment from financial news articles, they provide a more precise possible signal of market direction. Overall, LLMs are some of the most effective tools for stock prediction when integrating news sentiment into quantitative trading.
However, integrating LLM generated scores in prediction systems depends on time series models that that can handle long noisy sequences. Intraday stock data is extremely fine-grained and has fast transitions that require sensitivity to time and memory efficiency. Standard transformer models are often limited by the quadratic memory, Lack of Inductive Bias for Temporal Order and Memory Inefficiency. Therefore, Advanced architectures are needed to model such data effectively. 

Accordingly, we explore two advanced models for financial time series prediction: Reformer and Mamba. Reformer enhances transformer scalability through locality-sensitive hashing (LSH) and Invertible computational blocks, making attention for long sequences more efficient. Mamba presents a state-space modeling (SSM) approach that allows for sequence selectivity, while keeping linear complexity. Thus, Mamba is ideal for dense, high-resolution time series, such as prices as they arrive at a per minute resolution. 

In this research, we introduce a framework that merges LLM-generated sentiment scores with 1-minute interval stock price data of Apple Inc. (AAPL) over a 4-week time period. We developed a new dataset using ten different LLMs to retrieve a sentiment score for each news article and matched these with market data using timestamp matching and interpolation. Both Mamba and Reformer are trained separately across all ten LLM outcomes and performance is measured with Mean Squared Error (MSE) on a hold-out test set.

The main contributions of this paper are as follows:
\begin{itemize}
    \item Creation of a novel sentiment-integrated, minute-level financial dataset based on LLM outputs and intraday price changes. 
    \item Evaluation and Implementation of two efficient sequence models, Mamba and Reformer, for semantic-enhanced stock price prediction.

    \item Examining prediction precision over ten different LLM sentiment sources empirically.

    \item Key findings about all of the behaviors observed by the different LLMs and deep models in capturing market trends.

\end{itemize}
The remainder of this paper is organized as follows: Section 2 describes how we prepared our dataset and integrated sentiment; Section 3 describes the model architecture and training pipeline we created; Section 4 reports the experimental results and analysis; and finally, Section 5 gives a conclusion and proposes future work.

\section{Methodology}

\subsection{Data Collection}
This paper utilizes two datasets from the period April 4th at 4:01 PM EST to May 2nd at 4:00 PM EST of Apple Inc. (AAPL). The first dataset consists of financial news articles, while the second dataset contains high-frequency stock prices at 1-minute intervals during regular market hours. Both the datasets were later combined after applying sentiment scores to the news articles. The News API was used to fetch news headlines about Apple Inc. (AAPL) that were published between April 4, 2025, at 4:01 PM Eastern Time (ET), and May 2, 2025, at 4:00 PM ET. Each news article record contains the title and description, source, URL, and the timestamp when the article was published. To align with U.S. stock market hours, timestamps of the articles were converted to EST from UTC. Articles that were published outside regular business hours (before 9:30 AM or after 4:00 PM ET), as well as those from non-business days including weekends and Good Friday (April 18) were adjusted to the next trading day at 9:30 AM EST. High-frequency stock price data at 1-minute intervals during U.S. market hours was collected using the Polygon.io API. Each minute timestamp record includes the open, high, low, close prices and volume (OHLCV). Later, this dataset was used for aligning article timestamps with sentiment data and for computing technical indicators to support downstream predictive modeling.

\begin{figure}[htbp]
    \centering
    \includegraphics[width=\textwidth]{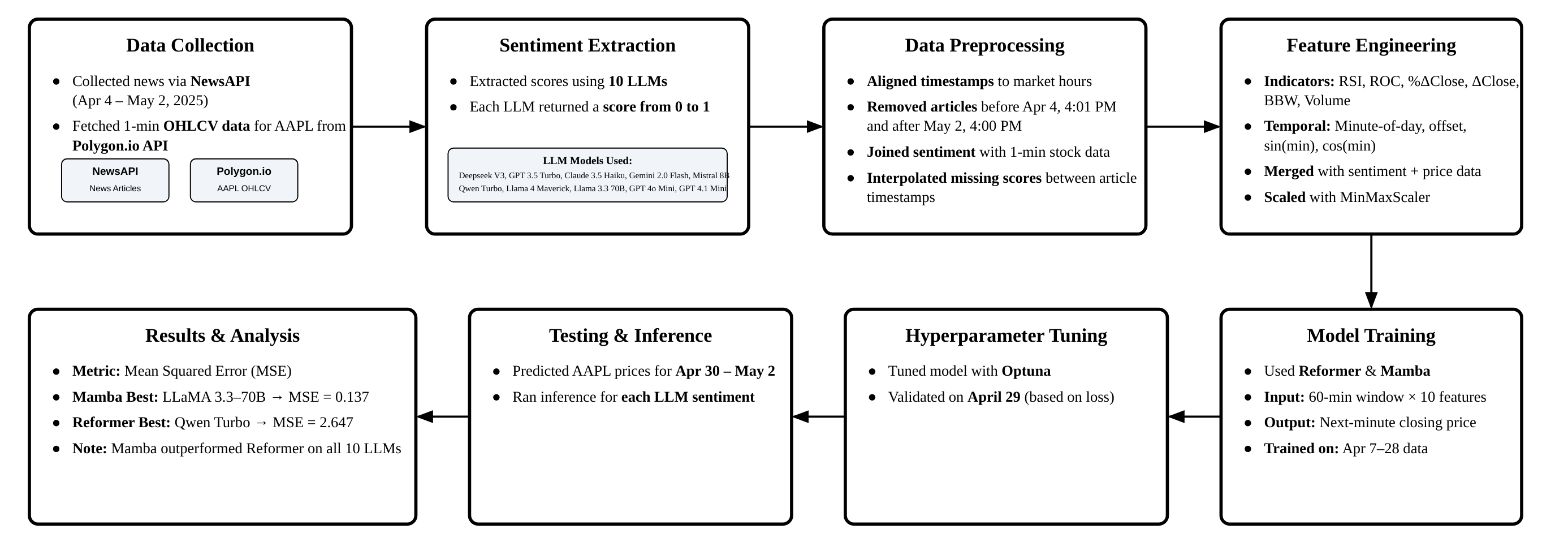}
    \caption{Overview of the LLM-Based Stock Price Prediction Pipeline.}
    \label{fig:pipeline-overview}
\end{figure}

\subsection{Sentiment Scores from LLMs}
To understand the impact of large language model generated sentiment on stock price prediction, sentiment scores were extracted from financial news articles using ten diverse large language models (LLMs) via APIs. 
Based on the news articles, each LLM generated a sentiment score ranging from 0 to 1, where 0 represents strongly negative sentiment, 1 represents strongly positive sentiment, and scores between 0.26 and 0.75 were considered as neutral.

 The sentiment analysis was performed using the following ten LLMs via API access: DeepSeek-V3, GPT-3.5 Turbo, LLaMA 3.3–70B, Claude 3.5 Haiku, GPT-4.1 Mini, GPT-4o Mini, LLaMA 4 Maverick, Gemini 2.0 Flash, Qwen Turbo, and Mistral ministral 8B. Each article was separately processed through all ten APIs, resulting in ten sentiment score columns per article. This enabled a direct comparison of each model’s influence on stock price predictions.
Sentiment scores were retrieved by accessing each LLM through its respective API platform. DeepSeek-V3 was accessed via the official DeepSeek API, while Qwen Turbo was obtained using the DashScope API provided by Alibaba Cloud. All other models including GPT-3.5 Turbo, GPT-4.1 Mini, GPT-4o Mini, Claude 3.5 Haiku, Gemini 2.0 Flash, LLaMA 3.3–70B, LLaMA 4 Maverick, and Mistral 8B were accessed using the OpenRouter API. Python scripts that fed news article titles to each model and stored the scores for integration with market data were used to fully automate the sentiment extraction process.

\subsection{Data Preprocessing and Normalization}

This process includes combining sentiment scores of articles with 1-minute intraday stock price data, handling missing values, performing interpolation.

Following timestamp alignment, as described in Section 5.1, each article’s sentiment score was mapped to the corresponding 1-minute interval in the stock data. Since all article timestamps were adjusted to valid NYSE trading hours, this enabled direct merging using the timestamp as a common key.

In instances where sentiment scores were unavailable for specific timestamps, linear interpolation was employed to estimate the values between two article-published sentiment scores.

\subsection{Feature Engineering}
To improve the model's predictive performance, a set of technical indicators are used. They are:
\begin{itemize}
    \item Relative Strength Index (RSI): determines whether the market is oversold by measuring the momentum of recent prices.
    \item Rate of Change (ROC): for a specific interval of time, it captures the velocity of price movement.
    \item Percentage Change in Closing Price ($\%\Delta$ Close): represents trends in short-term returns based on minute-level variations.
    \item Price Difference ($\Delta$Close): Represents the absolute difference between consecutive close prices.
    \item Bollinger Band Width (BBW): Quantifies market volatility by measuring the spread between the upper and lower Bollinger Bands.
    \item Trading Volume: finds unusual trading activity that could indicate changes in the mood of the market.
    \item Minute of the Day: indicates each timestamp's position within the daily trading session.
    \item Minute Offset: shows how many minutes have passed since the market opened.
    \item min\_sin and min\_cos: The models can learn intraday seasonality and periodic structures by encoding cyclical time patterns using sine and cosine transformations of the minute index.
    
\end{itemize}

\subsection{Final Dataset Structure}

Each row in the dataset represents a single 1-minute interval and contains:
\begin{itemize}
    \item The timestamp
    \item The stock’s close price
    \item Sentiment scores from ten LLMs
    \item Technical indicators, including RSI, ROC, $\Delta$Close, \%$\Delta$Close, BBW, and volume
    \item Temporal encodings, such as minute\_sin,minute\_cos,minute\_of\_day and
minute\_offset.
    
\end{itemize}

\section{Modeling and Evaluation}
This methodology part outlines the modeling strategy, training setup, hyperparameter tuning, and evaluation procedure for predicting stock prices using sentiment scores from large language models (LLMs). Two state-of-the-art models, Mamba and Reformer, were used separately for each LLM’s sentiment score.

\subsection{Model Overview}
This two advanced deep learning architectures, Mamba and Reformer are employed in this research to predict 1-minute interval stock prices for Apple Inc. (AAPL) based on historical prices, sentiment scores from 10 large language models (LLMs), and technical features.

\textbf{Mamba Forecast Model:}
The Mamba is initially based on the idea of state space model. The architecture comprises: 
\begin{itemize}
    \item Input Projection Layer: The input vectors were transformed by the fully connected
 layer into a 128-dimensional embedding space.
    \item Mamba Block: A single Mamba layer with d\_model=128, d\_state=16, d\_conv=4, and expand=2. This effectively learns long-range dependencies with low computational cost. 
    \item Output Layer: A fully connected layer predicts the next-minute closing price by transforming the sequence output into a scalar value. 
\end{itemize}

Optuna was used to tune key hyperparameters such as learning rate, weight decay, and batch size. The best parameters discovered during the validation phase were used to train the final model over 5 epochs.

\textbf{Reformer Forecast Model:}
The Reformer model is a Transformer variant that supports long-sequence learning with minimal memory usage using LSH based attention. Its architecture includes:
\begin{itemize}
    \item Input Projection Layer: converts each 10-dimensional feature vector to a Latent space (model\_dim=256).
    \item Reformer Block: Built from 3 layers (depth=3) with 8 attention heads (heads=8), a bucket size of 64, and n\_hashes=1. This enables the model to maintain efficiency while handling long sequences. 
    \item Output Layer: A linear layer generating the final scalar prediction.
\end{itemize}

This architecture makes sure that the sequence length is divisible by twice the bucket size (i.e., 128) by applying padding. Like the Mamba model, key parameters such as dropout rates and weight decay were tuned using Optuna.
Both models were implemented using Google Colab with CUDA-enabled T4 High-RAM GPUs after being initialized with Xavier initialization. Individual models were trained for each LLM sentiment score, allowing the architectures to learn the distinct influence of each LLM on stock price patterns.

\textbf{Input Format:} Both models accept input sequences consisting of 60 consecutive minutes. Each minute is represented by a vector of 10 features—comprising 9 engineered indicators (including technical and temporal attributes) and 1 sentiment score extracted from a specific large language model (LLM). The resulting input tensor has a shape of 60×10 for each training sample.

\textbf{Output Format:} The output of each model is computed through a fully connected (dense) projection layer that creates a single scalar value to predict the closing price for the subsequent minute.

\subsection{Mamba-Based Forecasting Framework}

The Mamba architecture, which is derived from Selective State Space Models (SSMs), consists of a streamlined yet powerful pipeline tuned for minute-level temporal signals and long-range dependencies. 
\begin{figure}
    \centering
    \includegraphics[width=1\linewidth]{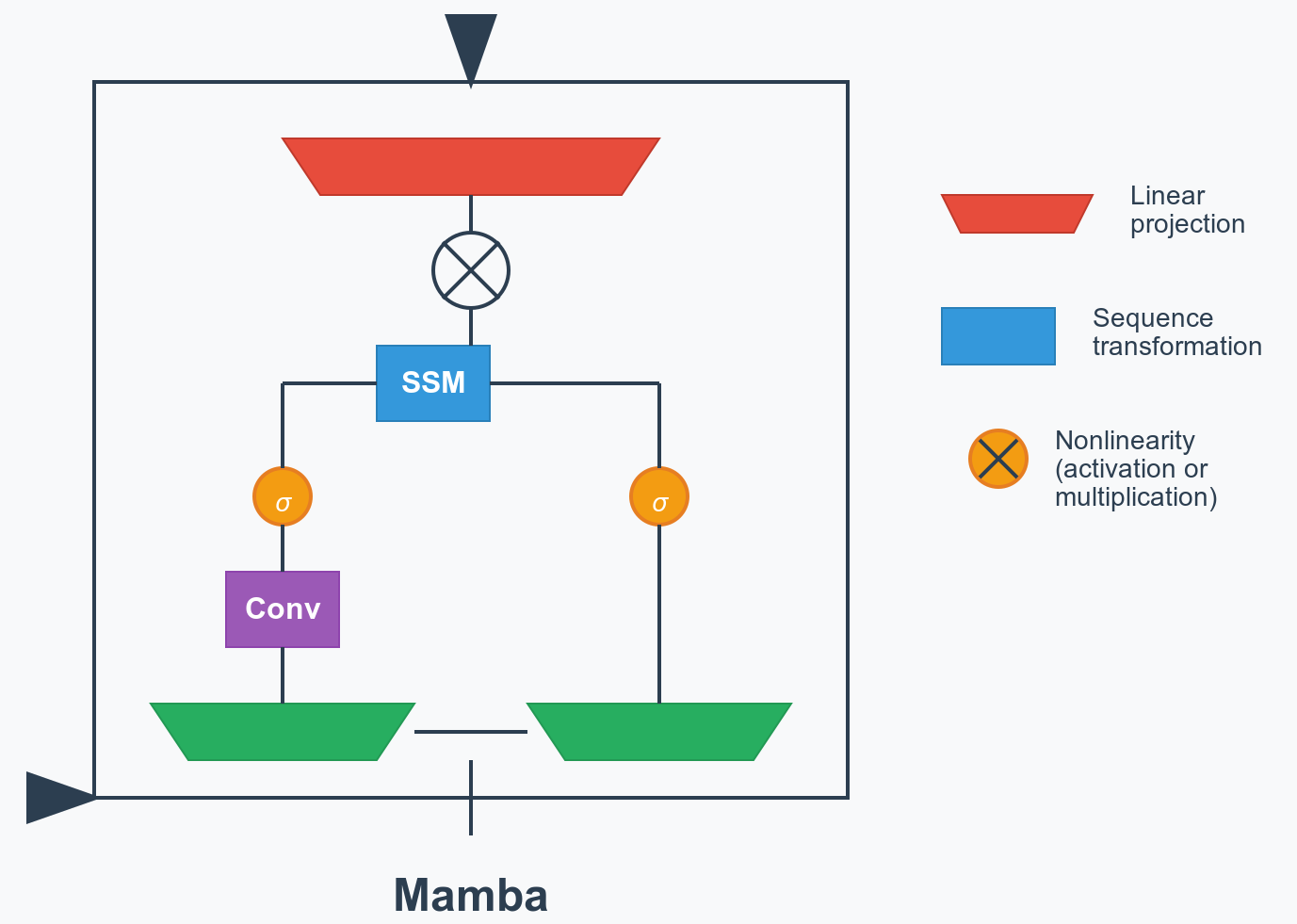}
    \caption{Architecture of the Mamba Block}
    \label{fig:enter-label}
\end{figure}
\textbf{Design rationale and architecture: } The architecture consists of a linear projection layer to map 10 input variables including technical indicators and sentiment into a 128-dimensional embedding space. This is followed by one Mamba block with \texttt{d\_model=128}, \texttt{d\_state=16}, \texttt{d\_conv=4}, and \texttt{expand=2}. The block effectively captures sequential dependencies and reduces short-term noise. The output layer emits a scalar prediction of the next-minute closing price.

\textbf{Engineering and training workflow: } 60-minute input windows were used for AAPL stock price data with technical indicators and 1 LLM-derived sentiment score in each model execution. MinMaxScaler was used to scale all input variables and targets while keeping the original price values for testing. Xavier initialization was applied for all trainable layers to improve model stability throughout training. The Models were optimized using AdamW optimizer, hyperparameter tuning being optimized by Optuna over 20 iterations. The last model was trained for over 5 epochs, and April 30 – May 2 data was tested with inverse-transformed predicted prices.

\textbf{Performance and robustness: } By filling the entire input sequence using the current LLM sentiment score, the model effectively integrates stationary technical patterns with volatile semantic inputs. Across multiple LLMs, Mamba demonstrated consistently low MSE values, performing best in scenarios with smoother sentiment gradients.

\subsection{Reformer-Based Forecasting Framework}

The Reformer architecture introduces a Transformer-like attention mechanism tailored for long-range stock market time series. It replaces traditional quadratic attention with Locality-Sensitive Hashing (LSH) and employs reversible residual layers, significantly lowering memory usage without compromising performance.

\textbf{Model Configuration and Mechanics:} \
Each \(60 \times 10\) input sequence is passed through a dense (fully connected) layer that projects it into a 256-dimensional latent space. A Reformer module consisting of three layers, each with 8 attention heads, processes the sequence using a bucket size of 64 and applies a single hash function (\(n\_hashes = 1\)). To maintain compatibility with the Reformer’s attention mechanism, padding is applied such that the total sequence length is divisible by \(2 \times \text{bucket\_size}\). The final output prediction is generated by applying a linear projection to the representation of the last time step in the sequence.

\textbf{Optimization and Training Strategy:} 
Each architecture was separately trained for each LLM-based sentiment score using Optuna for hyperparameter tuning (learning rate, dropout, weight decay, and batch size). 
Input variables were normalized using MinMaxScaler, and sentiment values were broadcast across the full 60-minute window. 
Training was executed for 5 epochs using the Adam optimizer, and predictions were inverse-transformed to restore their original scale. 
Dropout was applied in the attention layers to reduce overfitting, especially during isolated sentiment events.
\begin{figure}
    \centering
    \includegraphics[width=1\linewidth]{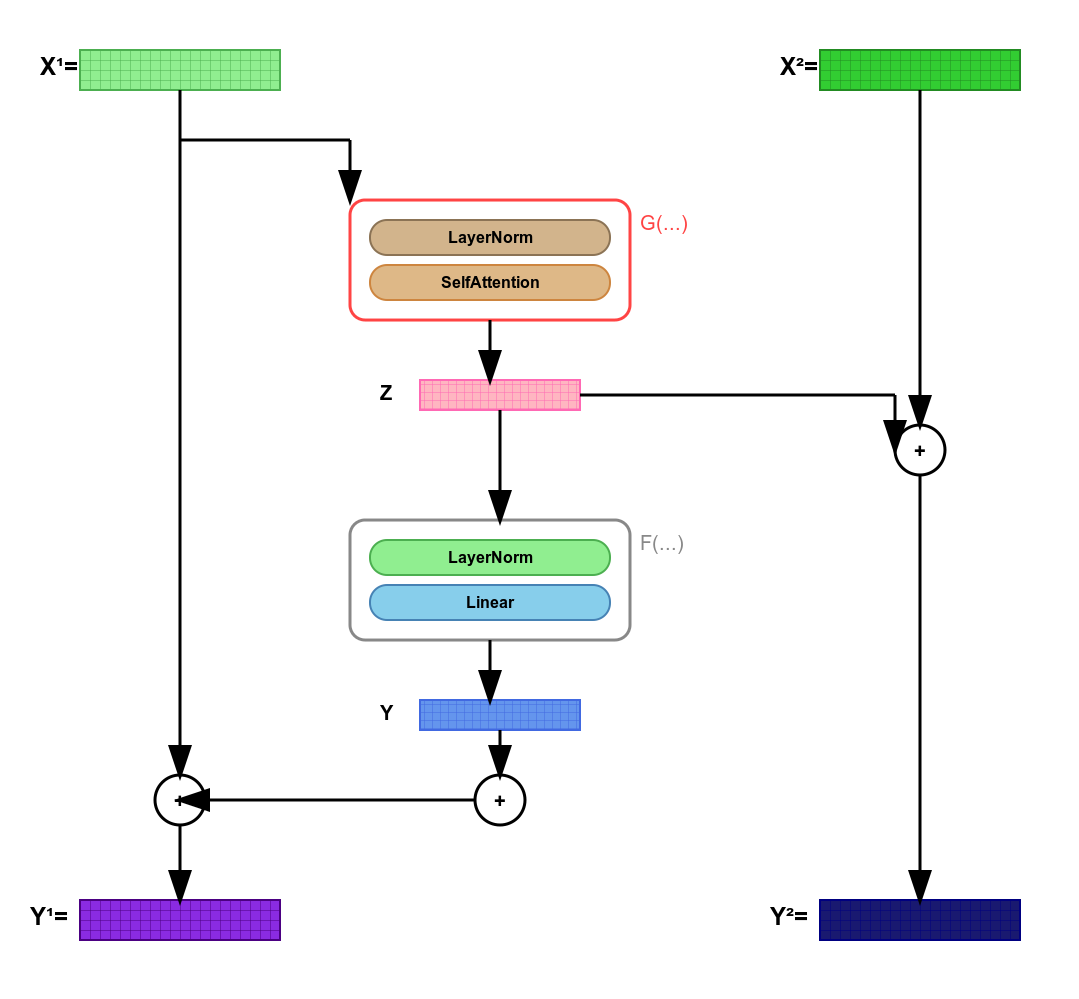}
    \caption{Overview of the Reformer architecture}
    \label{fig:enter-label}
\end{figure}
\textbf{Evaluation Insights:} 
The Reformer demonstrated strong capability in modeling feature interactions, particularly during periods of high market volatility and sentiment-driven trend shifts. Despite having a slightly higher computational cost than Mamba, it effectively captured both temporal and semantic dependencies essential for navigating complex market conditions.

\subsection{Data Splits}
Due to the dynamic nature of stock markets, data was split to minimize overfitting and improve predictive accuracy in the chronological order; each day we have 24*60 minutes sample price data,
\begin{itemize}
    \item Training Set: April 7 to April 28, 2025 (15 business days)
    \item Validation Set: April 29, 2025 (1 business day)
    \item Test Set: April 30 to May 2, 2025 (3 business days)
\end{itemize}

Every market day contains 390 1-minute entries from 9:30 AM to 4:00 PM EST. The 16-day training set allows it to learn relationships between stock characteristics and future price movements. while April 29 is used as the validation set for Optuna-based hyperparameter tuning. The final 3 days were used for testing, to generalize well on new data and avoid overfitting.

\subsection{Feature Scaling and Normalization}
The feature values were normalized using MinMaxScaler between 0 and 1 to avoid baised learning. For each training sample, the input vector consisted of 60 sequential 1-minute vectors, where each vector included 9 engineered features and 1 sentiment score extracted from a specific language model. 
The 9 engineered features were reshaped into a 2-D matrix from 3D tensor format. Individual scaling of the features was performed via MinMaxScaler from scikit-learn to transform values. The sentiment score which was unscaled resulting in the final input tensor of shape 60 × 10 for each sample.
During model training, Min-Max scaling was implemented to the next minute closing price . During the inference phase, estimated values were denormalized using the scaler derived from the training targets to recover the actual prices.

\subsection{Training Strategy}
The Mamba and Reformer models were trained to predict the next 1-minute closing price using historical data and sentiment scores generated by LLM. The training approach was designed across the market time series. Every input vector was generated from a 60-minute input window, forecasting the closing price at the 61st minute. This sliding window approach extracts local temporal trends and allows the model to learn patterns that reflect intraday market movements. Five iterations were used to tune the model to optimize a balance between computation speed and convergence. The difference between predicted and actual closing prices was minimized using the mean squared error loss function (MSE). The AdamW optimizer was employed to update weights with weight decay regularization to prevent overfitting. Xavier Uniform Initialization was used to intialize the weights for all linear and convolutional layers to promote stable gradients. Each sentiment score derived from ten different large language models (LLMs), was utilized independently by training a separate model instance for each language model. This method enabled the architecture to capture the unique influence of each LLM’s sentiment characteristics. Moreover, it supported a fair and consistent comparison of the predictive performance among different LLMs. The Google colab platform using CUDA-enabled T4 High-RAM GPUs was used to execute the training.

\subsection{Hyperparameter Tuning}
Hyperparameter optimization was executed using Optuna to boost the predictive performance of both the models. The primary objective was to determine the optimal parameter configurations that minimized validation error. The tuning process was based on a loss function that aimed to minimize the Mean Squared Error (MSE) on a fixed validation set dated April 29, 2025. This strategy enabled that the selected hyperparameters improved effective generalization on new and unseen data. The following hyperparameters were explored during the tuning process: 

\begin{itemize}
  \item \textbf{Mamba}
    \begin{itemize}
      \item Learning rate: log-uniform in the range \(1 \times 10^{-6}\) to \(1 \times 10^{-5}\)
      \item Batch size: \(\{32,\,64,\,128\}\)
      \item Weight decay: log-uniform in the range \(1 \times 10^{-6}\) to \(1 \times 10^{-5}\)
    \end{itemize}
  \item \textbf{Reformer}
    \begin{itemize}
      \item Learning rate: log-uniform in the range \(1 \times 10^{-6}\) to \(1 \times 10^{-5}\)
      \item Batch size: \(\{32,\,64\}\)
      \item Weight decay: log-uniform in the range \(1 \times 10^{-6}\) to \(1 \times 10^{-5}\)
      \item LSH dropout rate: uniform in the range \(0.10\) to \(0.25\)
    \end{itemize}
\end{itemize}

Hyperparameter tuning was performed using Optuna’s Tree-structured Parzen Estimator (TPE) sampler. For every model architecture and each sentiment input from individual LLMs, 20 independent trials were executed. The trial yielding the minimum MSE on the validation set was chosen as the best setup. Following the hyperparameter tuning, the best parameter configuration was used to retrain each model on the entire training set. The models were also trained for 5 iterations, which would not overfit the training set. This systematic and model-specific tuning method led to dramatic improvements in prediction performance and led each model to learn to project a distinct data distribution onto the sentiment scores for every LLM.

\subsection{Inference and Prediction}
The inference phase was carried out after finishing the training phase, utilizing the optimized hyperparameters to forecast stock closing prices at minute intervals for the testing period from 30~April to 2~May~2025. Each forecasting model was independently evaluated on the test set across the ten LLM-derived sentiment scores. For every forecast, a sliding input window of $60$ minutes was used to predict the closing price of the next minute. Each input vector consisted of nine engineered features merged with one LLM-specific sentiment score, resulting in a tensor of shape $60 \times 10$ for each prediction step. The model used historical data from time steps $t-60$ to $t-1$ (inclusive) to forecast the closing price at time $t$. This successive procedure was applied across all $390$ minutes per trading day over the three-day test duration, yielding a total of $1170$ forecasts per LLM per model. The predicted results were \emph{de-normalized} with the Min–Max scaler (fit on the training targets) to recover actual market prices, and each prediction was saved with its corresponding time label.



\section{Results and Discussion}

The Results part provides a comparative analysis for both Mamba and Reformer models for targeted AAPL’s minute-level closing prices. Each model was evaluated across ten separate LLM-derived sentiment score columns using both graphical trend comparison and MSE scores.

\subsection{Visual Evaluation}

\noindent\textbf{Mamba Model:} 
The sentiment scores derived from DeepSeek V3, Claude 3.5 Haiku, and GPT-4.1 Mini enabled strong correlation with actual prices, particularly during periods of high volatility. LLaMA 3.3 70B produced consistently accurate forecasts, effectively tracking trend shifts. In contrast, Mistral 8B and Gemini 2.0 Flash exhibited smoother curves but failed to respond sharply for changing trends.
\begin{figure}
    \centering
    \includegraphics[width=1\linewidth]{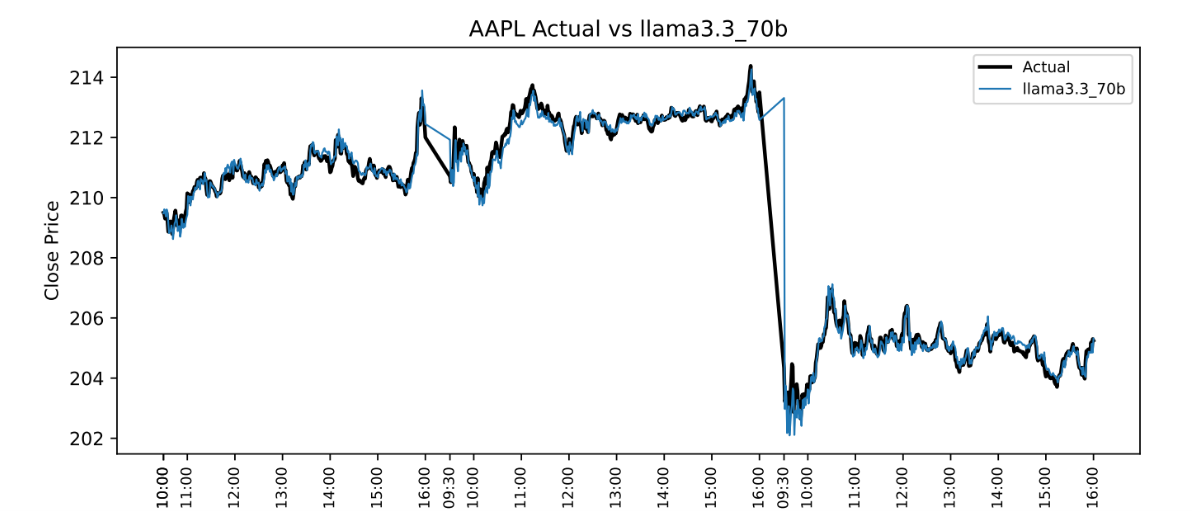}
    \caption{Mamba Model: AAPL minute-level prediction using LLaMA 3.3 70B sentiment scores}
    \label{fig:enter-label}
\end{figure}
\begin{figure}[htbp]
    \centering
    \includegraphics[width=1\linewidth]{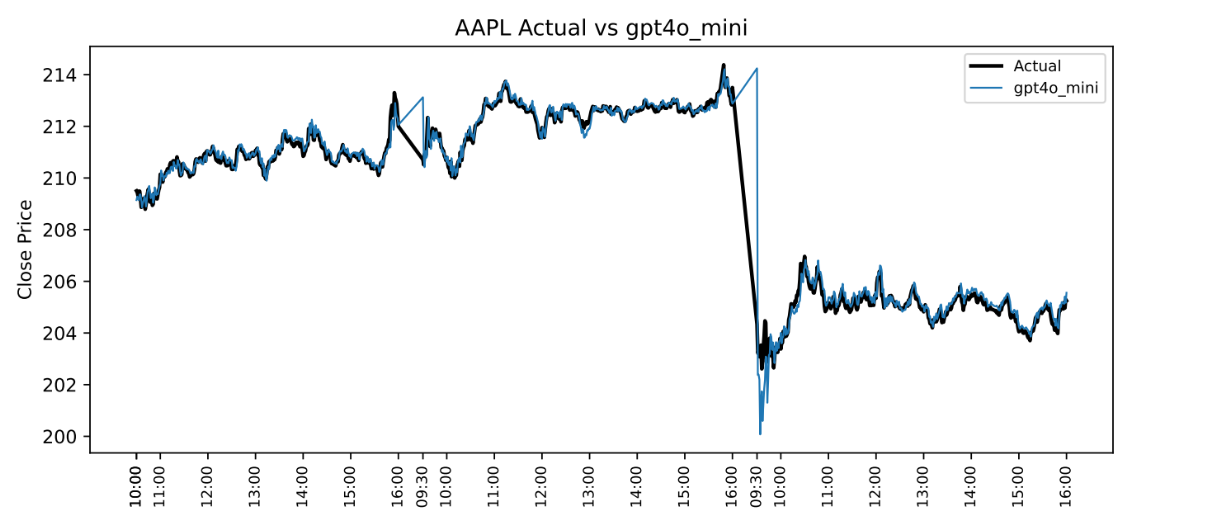}
    \caption{Mamba Model: AAPL minute-level prediction using GPT-4o Mini sentiment scores}
    \label{fig:mamba_dual}
\end{figure}

\vspace{0.5em}

\begin{figure}
    \centering
    \includegraphics[width=1\linewidth]{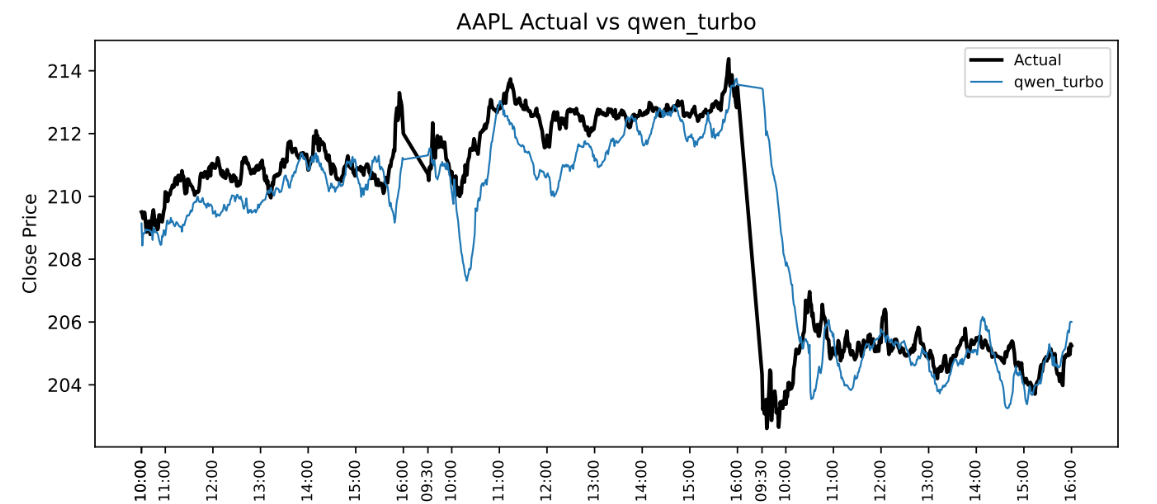}
    \caption{Reformer Model: AAPL minute-level prediction using Qwen Turbo sentiment scores}
    \label{fig:reformer_qwen}
\end{figure}
\noindent\textbf{Reformer Model:} 
Qwen Turbo and DeepSeek V3 are closely tracking actual market behavior, producing highly consistent predictions with only slight delays. GPT-3.5 Turbo, Claude 3.5 Haiku, and GPT-4.1 Mini attained reasonably accurate forecasts, though they displayed some delayed responsiveness to market shifts. LLaMA 4 Maverick and Gemini 2.0 Flash consistently underperformed by failing to respond to rapid market trends, resulting in smoother curves and reduced correlation with actual trends.



\subsection{Quantitative Evaluation (MSE)}       
The table below summarizes the Mean Squared Error (MSE) for both models across all LLM sentiment features. Lower values indicate more accurate forecasting.

\begin{table}[!ht]
    \centering
    \caption{ MSE metrics of both reformer and mamba.}\label{tab1}
    \begin{tabular}{|l|l|l|}
    \hline
        LLM Source & Reformer MSE & Mamba MSE \\ \hline
        Deepseek V3 & 2.9918 & 0.192 \\ \hline
        Qwen Turbo & \textbf{2.6468} & 0.2308 \\ \hline
        GPT 3.5 Turbo & 5.1086 & 0.4505 \\ \hline
        LLaMA 3.3 70B & 4.3236 & \textbf{0.1367} \\ \hline
        Claude 3.5 Haiku & 4.203 & 0.3554 \\ \hline
        GPT-4.1 Mini & 3.9706 & 0.2012 \\ \hline
        GPT-4o Mini & 3.2649 & 0.1859 \\ \hline
        Gemini 2.0 Flash & 7.2459 & 0.5047 \\ \hline
        LLaMA 4 Maverick & 4.2261 & 0.3324 \\ \hline
        Mistral ministral 7B & 3.018 & 0.3121 \\ \hline
    \end{tabular}
\end{table}

\begin{figure}
    \centering
    \includegraphics[width=1\linewidth]{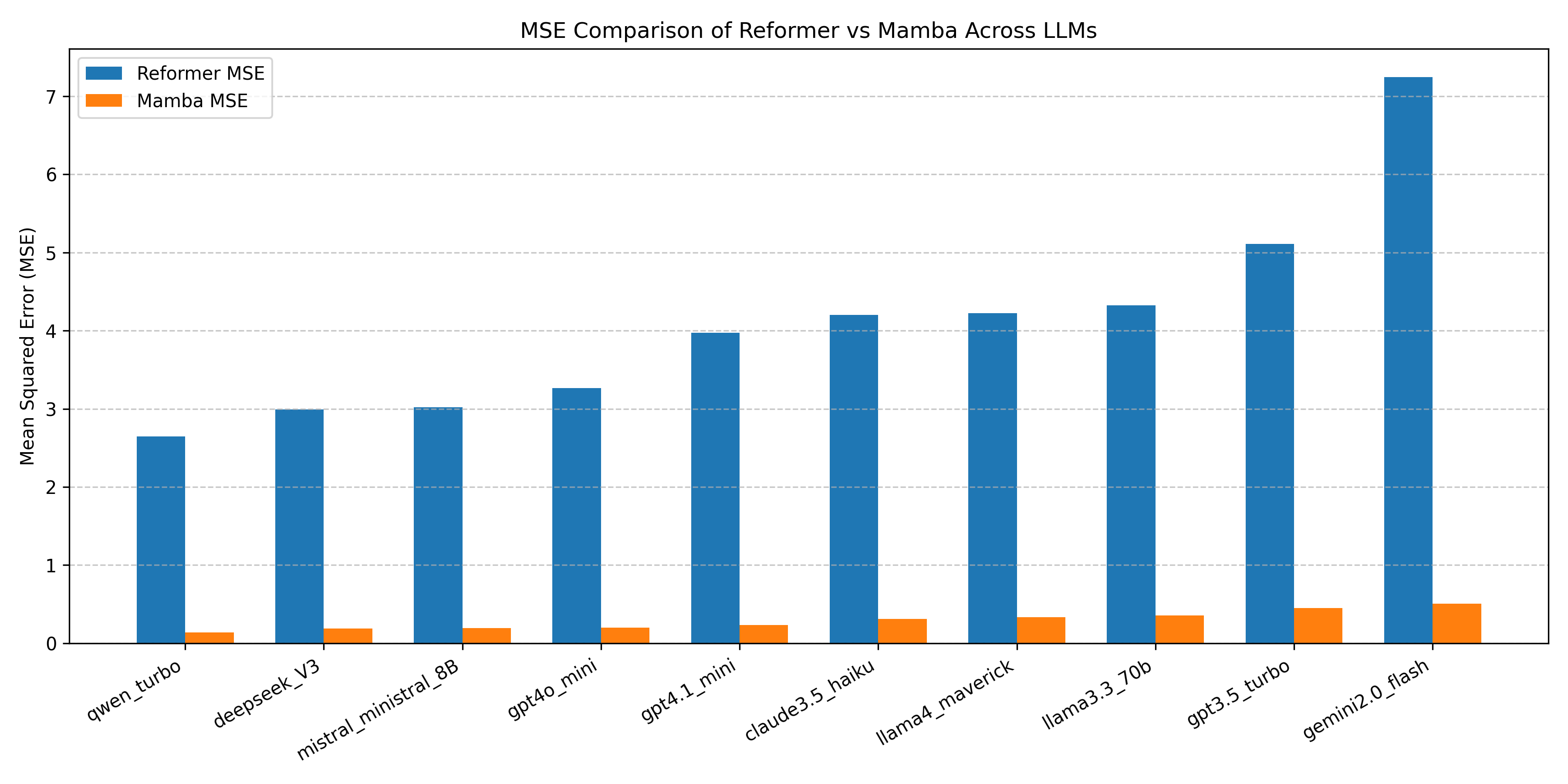}
    \caption{Mean Squared Error (MSE) comparison of Reformer and Mamba models across ten LLM-generated sentiment scores.}
    \label{fig:mse_comparison}
\end{figure}


\subsection{Key Insights}

\subsubsection{Best Performing Models}
\begin{itemize}
    \item \textbf{Mamba – Top Performer:} The Mamba model combined with LLaMA 3.3–70B had the lowest Mean Squared Error value of 0.137 across all LLM-based setups, showing accurate price tracking and high responsiveness to minute-level fluctuations. Mamba's structured state-space mechanism makes short-term predictions more accurate and allows for quick responses to changes in the market.
    
    \item \textbf{Reformer – Top Performer:} The Qwen Turbo setup had the best performance in the Reformer model, with an MSE value of 2.647. While it successfully tracked general trend directions, it was not as accurate as the best Mamba models.
\end{itemize}

\subsubsection{Comparative Evaluation}
\begin{itemize}
    \item \textbf{Performance Advantage – Mamba:} Mamba repeatedly outperformed Reformer across 9 out of 10 LLM-based inputs. Low-latency, sequential state representations are used in its architecture, which reduces lag and improves correlation with minute-level financial signals.
    
    \item \textbf{Trend Alignment – Reformer:} Despite higher error margins, the Reformer model showed reasonable outcomes when paired with Qwen Turbo and DeepSeek V3, capturing the broader structure of stock market trends. These setups showed strong trend correlation, though with reduced accuracy in detecting sharp price shifts.
\end{itemize}

\subsubsection{Model Behavior and Forecasting Sensitivity}
\begin{itemize}
    \item \textbf{Responsiveness vs. Smoothing Dynamics:} The Mamba model displayed greater responsiveness to minor market trends and price surges, maintaining tight alignment with ground truth data during rapid price shifts. On the other hand, Reformer leaned toward generalization, particularly when utilizing sentiment vectors from LLaMA 4 Maverick and Gemini 2.0 Flash, resulting in less sensitivity and smoother curves during rapid price fluctuations.
\end{itemize}

\section{Conclusion}
This research highlights the impact of combining LLM-based sentiment scores with real-time high-frequency stock price data for forecasting stock prices in real time. Through the use of the contextual understanding of semantic models with efficient deep learning architectures such as Reformer and Mamba, we are able to make considerable improvements in prediction precision.

Experimental results across ten different LLMs indicate that Mamba consistently achieves better short-term adaptability and lower MSEs, particularly when paired with LLaMA 3.3–70B and DeepSeek-V3. Additionally, while less precise in rapidly changing environments, Reformer was still able to capture directional trends accurately using architectures optimized for long sequences.

This study demonstrates that LLM-based sentiment can be a strong predictive signal when used in models that can capture both short- and long-term dependencies. Future research will include extending the time period and adding additional stocks, while also scaling the prompts of LLMs for texts related to finance.

In summary, this study presents a novel sentiment–temporal fusion pipeline and a performance comparison of cutting-edge deep learning models for intraday financial forecasting.

\appendix

\section*{Appendix: Sentiment Retrieval Process and Sample Scores}

\subsection*{Prompt and Code Example (LLaMA 3.3–70B)}

To retrieve sentiment scores from financial news article titles, a consistent prompting strategy was applied across all large language models (LLMs). Below is the prompt and Python code used for the \textbf{meta-llama/llama-3.3-70b-instruct} model via OpenRouter API. This serves as a representative example for all other LLMs used in this study.

\textbf{Prompt:}
\begin{quote}
\ttfamily
You are a sentiment analysis AI. Return only a numeric sentiment score between 0 and 1. Do not include any words, symbols, or extra characters—only the number. Use the 'Title' column from the dataset as input for sentiment analysis. Apply the given architecture (based on the meta-llama/llama-3.3-70b-instruct model via OpenRouter). Process the dataset using up to 50 parallel threads for efficiency. Save the sentiment scores in a CSV file alongside their corresponding titles.
\end{quote}

\textbf{Python Code Snippet:}
\begin{verbatim}
from openai import OpenAI

client = OpenAI(
  base_url="https://openrouter.ai/api/v1",
  api_key="<OPENROUTER_API_KEY>",
)

completion = client.chat.completions.create(
  extra_headers={
    "HTTP-Referer": "<YOUR_SITE_URL>",  # Optional: Your website URL for OpenRouter rankings
    "X-Title": "<YOUR_SITE_NAME>",      # Optional: Your site/app name
  },
  extra_body={},
  model="meta-llama/llama-3.3-70b-instruct",
  messages=[
    {
      "role": "user",
      "content": "What is the meaning of life?"
    }
  ]
)

print(completion.choices[0].message.content)
\end{verbatim}

\textit{Note:} A similar structure and prompt were adapted for the remaining nine LLMs used in the study. Only the model name in the API call was changed accordingly.

\clearpage

\vspace{1em}
\subsection*{Sample Sentiment Scores from 10 LLMs}

The following table presents sentiment scores (ranging from 0 to 1) for 20 Apple-related articles, evaluated using ten different large language models (LLMs). This data serves as a qualitative sample to illustrate inter-model consistency and variance.

\begin{table*}[!ht]
\centering
\caption{Sentiment Scores of 20 Apple Articles Across 10 LLMs}
\label{tab:sentiment-appendix}
\resizebox{\textwidth}{!}{%
\begin{tabular}{|p{5cm}|c|c|c|c|c|c|c|c|c|c|}
\hline
\textbf{Title} & \textbf{DeepSeek} & \textbf{GPT 3.5} & \textbf{Claude} & \textbf{Gemini} & \textbf{Qwen} & \textbf{LLaMA 3.3} & \textbf{LLaMA 4} & \textbf{GPT 4.1} & \textbf{GPT 4o} & \textbf{Mistral} \\ \hline
Reality gets in the way of Elon Musk’s latest misg & 0.30 & 0.20 & 0.20 & 0.10 & 0.20 & 0.20 & 0.20 & 0.20 & 0.10 & 0.20 \\ \hline
Why Webull Corporation (BULL) Went Up On Friday & 0.60 & 0.60 & 0.60 & 0.60 & 0.20 & 0.60 & 0.60 & 0.60 & 0.50 & 0.50 \\ \hline
Scaling 6 Products to \$100M+ ARR Each: Samsara’s C & 0.80 & 0.80 & 0.70 & 0.70 & 0.75 & 0.85 & 0.80 & 0.70 & 0.70 & 0.90 \\ \hline
‘Snow White:’ Lebanon Bans Disney Remake Over Incl & 0.80 & 0.80 & 0.70 & 0.70 & 0.75 & 0.85 & 0.80 & 0.70 & 0.70 & 0.80 \\ \hline
ESG Is Coming for Your Candy Bars & 0.30 & 0.20 & 0.30 & 0.30 & 0.20 & 0.20 & 0.20 & 0.30 & 0.20 & 0.20 \\ \hline
Apple Releases iOS 18.4.1 With Bug Fixes & 0.40 & 0.20 & 0.30 & 0.30 & 0.20 & 0.40 & 0.30 & 0.30 & 0.70 & 0.20 \\ \hline
Rugged Display Market Size to Worth USD 18.02 Bill & 0.70 & 0.80 & 0.50 & 0.60 & 0.20 & 0.60 & 0.60 & 0.70 & 0.50 & 0.70 \\ \hline
DOJ Files Amicus Brief Supporting Pro-Second Amend & 0.70 & 0.80 & 0.70 & 0.60 & 0.50 & 0.62 & 0.60 & 0.50 & 0.50 & 0.50 \\ \hline
Lyft’s Cofounder Stepped Down Two Years Ago. Now H & 0.60 & 0.80 & 0.50 & 0.60 & 0.30 & 0.67 & 0.60 & 0.70 & 0.70 & 0.80 \\ \hline
This independent testing lab confirms Saily users  & 0.50 & 0.50 & 0.50 & 0.50 & 0.50 & 0.50 & 0.50 & 0.50 & 0.50 & 0.50 \\ \hline
A small fashion brand owner took to TikTok to lay  & 0.80 & 0.90 & 0.70 & 0.80 & 0.85 & 0.80 & 0.90 & 0.90 & 0.80 & 0.80 \\ \hline
Argentina peso expected to hit lower band, says de & 0.30 & 0.30 & 0.30 & 0.20 & 0.25 & 0.20 & 0.40 & 0.40 & 0.20 & 0.20 \\ \hline
Our University’s Commitment to You & 0.40 & 0.20 & 0.30 & 0.30 & 0.25 & 0.40 & 0.30 & 0.40 & 0.30 & 0.20 \\ \hline
Why Aurora Innovation Inc. (AUR) Went Down On Frid & 0.80 & 0.80 & 0.50 & 0.60 & 0.50 & 0.80 & 0.70 & 0.80 & 0.70 & 0.90 \\ \hline
5 security features in Windows 11 you should activ & 0.30 & 0.20 & 0.40 & 0.20 & 0.20 & 0.20 & 0.30 & 0.40 & 0.20 & 0.20 \\ \hline
Alaska elite qualifying miles are now posting for  & 0.30 & 0.20 & 0.40 & 0.20 & 0.20 & 0.20 & 0.30 & 0.40 & 0.40 & 0.20 \\ \hline
Goldman Sachs Withdraws Recession Prediction After & 0.70 & 0.70 & 0.70 & 0.60 & 0.20 & 0.60 & 0.70 & 0.60 & 0.50 & 0.70 \\ \hline
Quantum Health Founder Makes Humanizing Healthcare & 0.70 & 0.80 & 0.60 & 0.70 & 0.50 & 0.70 & 0.70 & 0.60 & 0.50 & 0.70 \\ \hline
Sacai’s Nike Zegamadome SP Launches On May 9th & 0.70 & 0.80 & 0.70 & 0.60 & 0.50 & 0.62 & 0.60 & 0.60 & 0.70 & 0.70 \\ \hline
Which Boycotts Are Ongoing Across the US? & 0.70 & 0.80 & 0.70 & 0.60 & 0.65 & 0.62 & 0.60 & 0.60 & 0.70 & 0.70 \\ \hline
\end{tabular}
}
\end{table*}

%
%
%
%

\end{document}